# Direct Observation of the Skyrmion Hall Effect

Wanjun Jiang[1,*,†], Xichao Zhang[2,*], Guoqiang Yu[3], Wei Zhang[1],

M. Benjamin Jungfleisch[1], John E. Pearson[1], Olle Heinonen[1,4], Kang L. Wang[3],

Yan Zhou[2], Axel Hoffmann[1,†], Suzanne G. E. te Velthuis[1,†]

[1]Materials Science Division, Argonne National Laboratory, Lemont, Illinois, USA, 60439

[2]Department of Physics, The University of Hong Kong, Hong Kong, China

[3]Department of Electric Engineering, University of California, Los Angeles,

California, USA, 90095

[4]Northwestern-Argonne Institute of Science and Engineering,

Northwestern University, Evanston, IL 60208

[*]These authors contributed equally.

[†]To whom correspondence should be addressed.

E-mail: jiangw@anl.gov, hoffmann@anl.gov, tevelthuis@anl.gov

**The well-known Hall effect describes the transverse deflection of charged particles (electrons/holes) in an electric-current carrying conductor under the influence of perpendicular magnetic fields, as a result of the Lorentz force. Similarly, it is intriguing to examine if quasi-particles without an electric charge, but with a topological charge[1-4], show related transverse motion. Chiral magnetic skyrmions with a well-defined spin topology resulting in a unit topological charge serve as good candidates to test this hypothesis[1-3,5-11]. In spite of the recent progress made on investigating magnetic skyrmions[2,4,6-8,12-19], direct observation of the skyrmion Hall effect in real space has, remained elusive. Here, by using a current-induced spin Hall spin torque[13,20-23], we experimentally observe the skyrmion Hall effect by driving skyrmions from creep motion into the steady flow motion regime. We observe a Hall angle for the magnetic skyrmion motion as large as $15°$ for current densities smaller than $10^7 \text{ A/cm}^2$ at room temperature. The experimental observation of transverse transport of skyrmions due to topological charge may potentially create many exciting opportunities for the emerging field of skyrmionics, including novel applications such as topological selection.**

Because of their topologically non-trivial spin textures, chiral magnetic skyrmions enable many intriguing phenomena based on their topology[2-4], such as emergent electro-dynamics[10] and effective magnetic monopoles[11]. As compared to most (vortex-like) Bloch skyrmions in bulk chiral materials[2,5,9], utilizing interfacial inversion symmetry breaking[24] in heavy metal/ultrathin ferromagnet/insulator hetero-structures has enabled the observation of robust chiral spin textures such as (hedgehog) Néel skyrmions, even at room temperature[6,8,13,16-18,25]. In addition, the giant spin-orbit torques from the spin Hall effect of the incorporated heavy metals (typically Ta, Pt, and W)[21,22], provides energetically efficient avenues for the electrical generation and manipulation of magnetic skyrmions[13,18,26].

By considering an isolated Néel skyrmion as a rigid point-like particle [with its spin texture illustrated in Fig. 1 (a)], the translational motion driven by the spin Hall effect can be described by a modified Thiele equation[2,3,27-29]:

$$\boldsymbol{G} \times \boldsymbol{v} - \alpha \boldsymbol{D} \cdot \boldsymbol{v} + 4\pi \overleftrightarrow{\boldsymbol{B}} \cdot J_{hm} = \boldsymbol{0} \qquad (1)$$

Here $\boldsymbol{G} = (0, 0, -4\pi Q)$ is the gyromagnetic coupling vector with the topological charge $Q$ being defined as $Q = {}^{1}\!/\!_{4\pi} \int \boldsymbol{m} \cdot (\partial_x \boldsymbol{m} \times \partial_y \boldsymbol{m})\, dxdy$. $\boldsymbol{v} = (v_x, v_y)$ is the skyrmion drift velocity along the $x$ and $y$ axes, respectively. $\alpha$ is the magnetic damping coefficient, and $\boldsymbol{D}$ is the dissipative force tensor. The tensor $\overleftrightarrow{\boldsymbol{B}}$ quantifies the efficiency of the spin Hall spin torque over the 2-dimensional spin texture of the skyrmion, $J_{hm} = J_s/\theta_{sh}$ is the electrical current density flowing in the heavy metal, $J_s$ is the spin current density, $\theta_{sh}$ is the spin Hall angle of the heavy metal. The first term in eq. (1) is the topological Magnus force that results in the transverse motion of skyrmions with respect to the driving current[3,7,28,29]. This term acts equivalently to the Lorentz force for electric charge, and thus gives rise to a Hall-like behaviour of magnetic skyrmions[4]. The second term is the dissipative force that is linked to the intrinsic magnetic damping of a moving magnetic skyrmion, and the third term is the driving force from the spin Hall spin torque. We note that Eq. (1) does not include the possible pinning effects that may impede skyrmion motion due to the presence of materials imperfections, nor does it include the possibility of exciting internal degrees of freedom of the skyrmion. Such internal degrees of freedom may modify the dynamics and dissipation of the driven skyrmion.

Upon applying a spatially homogeneous current along the $x$ direction $J_{hm} = (j_x, 0)$, the resultant velocity along the $x$ and $y$ axes can be calculated as $v_x = \frac{\alpha D}{1+\alpha^2 D^2} B_0 j_x$, $v_y = \frac{1}{1+\alpha^2 D^2} B_0 j_x$, respectively[28,29]. Here $B_0$ is a constant that can be estimated based on the detailed spin configuration. This leads to the following expression for the ratio of in-plane velocity components to be written as:

$$\frac{v_y}{v_x} = \frac{1}{\alpha D} \quad (2)$$

where $\alpha \approx 0.02$ is the damping parameter (for the magnetic bilayer involved in this study). As shown in the supplementary material $D = \pi^2 d / 8\gamma_{dw}$ with $d$ being the skyrmion diameter and $\gamma_{dw}$ being the width of chiral domain wall. This means that $D$ varies from about 2 for 10 nm sized skyrmion (assuming $D = 2\gamma_{dw}$) to about 100 for µm sized skyrmions with a narrow domain wall of 10 nm width. Thus for small Néel skyrmions $v_y/v_x \gg 1$, results in mostly perpendicular motion, while for larger skyrmions the perpendicular motion is less pronounced with $v_y/v_x \approx 1$. In either case, this should lead to the accumulation of skyrmions at the device edge, as demonstrated by micromagnetic simulations shown in the Fig. 1 (b). The size of the numerically simulated skyrmions in Fig. 1(b) is about 100 nm, which results in $D \approx 12$. Thus for $\alpha \approx 0.02$, resulting in $v_y/v_x \approx 4$, the skyrmion Hall angle corresponds to about $\Phi_{sk} \approx 76°$.

Note that during the motion along the edge, the size of skyrmion shrinks slightly due to the repulsive force from the edge[12,25,28,29].

Reversing the perpendicular magnetic fields, and correspondingly the magnetization from positive to negative, the sign of the skyrmion topological charge is also reversed from $Q = -1$ in the upper panel to $Q = +1$ to the lower panel of Fig. 1 (d). This is because the topological charge [$Q = 1/4\pi \int \bm{m} \cdot (\partial_x \bm{m} \times \partial_y \bm{m})\, dxdy$] is an odd function of the magnetization vector $\bm{m}$, and therefore reverses sign upon inversion of the spin textures, *i.e.*, by applying opposite magnetic fields. This sign reversal leads to opposite direction of the topological Magnus force (since $\bm{G} \times \bm{v}$ is linked to the sign of topological charge), and hence the accumulation of skyrmions at the opposite edge. This behavior resembles phenomenologically the electronic Hall effect of electrons/holes in conductors in the presence of perpendicular magnetic fields, Fig 1 (c); therefore this behavior is referred to as the *skyrmion Hall effect*, Fig. 1 (d). While the occurrence of the skyrmion Hall effect has been predicted theoretically, and suggested numerically via micromagnetic simulations[4,28,29], an experimental observation has still been lacking and is addressed here.

By utilizing a geometrical constriction, we have previously demonstrated that spatially divergent spin Hall spin torques can dynamically convert chiral band domain into Néel skyrmions[13]. That work was performed with an interfacially asymmetric Ta(5nm)/Co$_{20}$Fe$_{60}$B$_{20}$(CoFeB)(1.1nm)/TaO$_x$(3nm) trilayer with an interfacial

Dzyaloshinskii-Moriya interaction[30] of strength ≤0.5 $mJ/m^2$. These electrically generated skyrmions did, however, not show any clear signature of the skyrmion Hall effect[13], in contrast to expectations from theoretical modeling[26]. As these experiments were performed at low current densities ($j_e < 1 \times 10^5$ $A/cm^2$), the absence of transverse motion can be attributed to the creep motion of skyrmions in the low current density regime, in which the direction of motion was influenced strongly by the pining potential of randomly distributed defects[31-33]. By progressively increasing the current density, it should be possible to drive skyrmions from the creep motion regime into the steady flow motion regime, as suggested by recent theoretical studies on the collective transport of skyrmions with random disorders/defects[31-33].

In the present study, we employ a modified design, a Hall bar devices of the same material system Ta/CoFeB/TaO$_x$ with dimensions 100 (width) × 500 (length) μm$^2$, wherein a larger current density can be applied. We note that a few skyrmions were observed upon sweeping perpendicular magnetic fields. Current-driven imaging data were acquired by using a polar magneto-optical Kerr effect (MOKE) microscope in a differential mode at room temperature.

We first discuss the creep motion of the skyrmions in the low current densities regime. Individual polar-MOKE images are shown in Figs. 2 (a) – 2 (f) for $Q = -1$ skyrmions and in Figs. 2 (g) – 2 (l) for $Q = +1$ skyrmions, respectively. Note that we studied a single skyrmion that is isolated from other skyrmions to avoid complications due to skyrmion-skyrmion interactions. These two experiments were performed by applying pulsed electron currents of amplitude $j_e = +1.3 \times 10^6$ A/cm² and with a duration of 50 $\mu s$. In this report, $j_e$ denotes the electron motion; i.e., opposite to the charge motion. The red arrow refers to a positive electron motion direction ($+j_e$) from left to right. The current density is normalized by the total thickness of Ta (5 nm) and CoFeB (1.1 nm). By comparing the trajectories shown in Figs. 2 (f) for the $Q = -1$ skyrmion and 2 (l) for the $Q = +1$ skyrmion, it is clear that the stochastic motion is observed without net transverse components.

By increasing the current density to $j_e = +2.8 \times 10^6$ A/cm², it is observed that the direction of motion develops a well-defined transverse component, which is exemplified by a straight and diagonal trajectory. Figs. 2 (m) – (r) correspond to a $Q = -1$ skyrmion, and Figs. 2 (s) – (x) to a $Q = +1$ skyrmion. The opposite sign of slopes in Figs. 2 (r) and 2 (x) are consistent with the opposite sign of the topological Magnus force, that gives rise to opposite directions for the transverse motion. By reversing the electron current direction, the direction of motion is also reversed, as shown in the supplementary information.

A current-velocity relationship is subsequently established and is shown in Fig. 3 (a), indicating a monotonic increase of the average velocity as a function of current density. The average velocity ($\bar{v}$) is defined as $\bar{v} = \mathcal{L}/(N \cdot \Delta t)$, where $\mathcal{L}$ is the total displacement, $N$ is the number of pulses, and $\Delta t$ is the duration of pulse. The number of pulses was typically chosen to be $N > 10$ to minimize the uncertainty due to the stochastic motion of skyrmion in the creep motion regime. For a fixed pulse duration of 50 $\mu s$, there is a threshold depinning current density $j_e = (0.6 \pm 0.1) \times 10^6$ A/cm$^2$, below which skyrmions remain stationary [light blue regime in Fig. 3 (a)]. Above this threshold depinning current two features were observed: (A) stochastic migration of skyrmions following the electron current direction when $j_e < 1.5 \times 10^6$ A/cm$^2$ [light orange regime in Fig. 3 (a)]. (B) motion of skyrmions with a well-defined transverse velocity when the current density $j_e > 1.5 \times 10^6$ A/cm$^2$ [light green regime in Fig. 3 (b)]. For example, the velocity is estimated to be $\bar{v} \approx 0.75 \pm 0.02$ m/s at a current density $j_e = 6.2 \times 10^6$ A/cm$^2$. It is also noticed that the threshold depinning current density evolves as a function of pulse duration. Namely, for shorter pulses, larger amplitudes are required to result in skyrmion motion, indicative of a thermally assisted depinning process of magnetic skyrmions from local pinning sites due to disorder.

The onset of transverse motion is shown in Fig. 3 (b). The deviation of skyrmion motion with respect to the applied current direction $(+x)$ can be quantified by a skyrmion Hall angle $\Phi_{sk} = tan^{-1}\left(v_y/v_x\right)$. With increasing current density ($j_e > 1.5 \times 10^6$ A/cm$^2$), the ratio of $v_y/v_x$ (dark red symbols) and consequently the skyrmion Hall angle $\Phi_{sk}$ (blue symbols) increase monotonically. The skyrmion Hall angle can be as large as $\Phi_{sk} \approx 15°$ with a ratio of $v_y/v_x \approx 0.28$ at the maximum current density $j_e = 6.2 \times 10^6$ A/cm$^2$ (limited by the present instrument). Given the roughly linear dependence on the current density without indication of saturation, it is expected that the skyrmion Hall angle $\Phi_{sk}$ can be even larger for even higher current densities.

This current density dependence is contradictory to the simple theoretical prediction given in equation (2), which suggests a constant value of $v_y/v_x$ that is simply dictated by $1/\alpha D$ and is independent of the driving current. On the other hand, the magnitude obtained at the highest achievable current density is consistent with what might be expected. While for most theoretical studies, small skyrmions are considered (10 nm), leading to a small dissipative term $D$ and $v_y/v_x > 1$, the skyrmions imaged here are significantly larger (> 1000 nm). For skyrmions of diameter $\approx$ 1000 nm with a fixed domain wall widths $\gamma_{dw} \approx 10$ nm, the value of the dissipative term is estimated to be $D \approx 120$ and increases proportionally with the area of the skyrmion, as discussed in Fig. S1 in the Supplementary Information. Subsequently, a constant ratio of $v_y/v_x \approx 0.4$ corresponding to a skyrmion Hall angle $\Phi_{sk} \approx 22°$ is estimated as an upper limit for our system, which is compatible with the observed values of $v_y/v_x$ being less than 0.3 for current density $j_e < 7 \times 10^6$ A/cm$^2$. Further increase of the current density (which is beyond the present instrument limit) could thus lead to higher values of $v_y/v_x$.

One possible reason for the apparent discrepancy between our experimental observation of current dependent increasing value of $v_y/v_x$ and the simple prediction of equation (2) is the presence of pinning that affects the skyrmion motion[31-33]. Such pinning may originate from random defects/disorder in the sputtered films, as suggested by our experimentally observed threshold depinning of skyrmions and stochastic motion at low driving currents. Detailed theoretical investigation of the dynamics of skyrmions interacting with randomly distributed disorders/defects has shown a significant reduction of the skyrmion Hall angle $\Phi_{sk}$, as well as complex skyrmion trajectories[31-33]. Specifically, the skyrmion Hall angle is minimized around the depinning threshold and increases monotonically with the driving current, due to the pronounced side-jump scattering between skyrmions and scattering potential. This is reminiscent of our experimental observations in the absence of interactions between multiple skyrmions. Experimentally for $j_e < 1.5 \times 10^6 \text{ A/cm}^2$, skyrmions escape from the pining potential and exhibiting a hopping-like motion along the driving direction with a zero skyrmion Hall angle. In the strong-driving regime $j_e > 1.5 \times 10^6 \text{ A/cm}^2$, increasing the driving force increases monotonically the skyrmion Hall angle $\Phi_{sk}$. Note that there should exist an upper limit, where the amplitude of $\Phi_{sk}$ is limited by the intrinsic damping and dissipation[28,29].

On the other hand, given the large size of magnetic skyrmions (of diameter ≈ 1000 nm), and the weak strength of interfacial DMI $D_{dmi} \leq 0.5 \text{ mJ/m}^2$ in the present material system, it is also possible that the current induced spin Hall spin torque could modify the spin texture of skyrmion[25,30]. This could invalidate the assumption of a rigid spin structure underlying the Thiele's equation. Indeed, micromagnetically generated large skyrmions in the weak DMI regime $(D_{dmi} \leq 0.5 \text{ mJ/m}^2)$ exhibit considerable dynamics because of internal degrees of freedom[30,34]. These could potentially modify the response to an external field as well as the response to a pinning potential, in addition to modifying how energy is dissipated in the system.

Furthermore, it is expected that close to the sample edge the motion of skyrmion is modified. The experimentally observed motion is shown in Fig. 4. The skyrmion marked by the green circle is annihilated by the structural defects, which has also been observed in other material systems[18]. In the higher current density regime, a $Q = +1$ skyrmion in the center of the device move with a fixed skyrmion Hall angle of $\Phi_{sk} \approx 8°$ and an average velocity of $\bar{v} \approx 0.25 \pm 0.01 \text{ m/s}$, as shown in Figs. 2 (s) - (x). However, a $Q = +1$ skyrmion close to the edge shows an "oscillatory" transport feature at the same current density. Namely, the moving skyrmion is repelled away from the edge, as a result of the competition between the driving force from the spin Hall spin torque and the repulsive force from the edge of the sample[12,28,29]. This repulsive force, which is of dipolar interaction in nature, kicks the skyrmion back from the edge after a pulse current. This motions is clearly visible on the relatively slow observational time scales, indicative of a slow, almost diffusive motion, probably due to creep because of random defects. This explains the "oscillatory" trajectory, and hence the absence of well-defined skyrmion Hall angle close to the edge, as summarized in Fig. 4 (g). The corresponding average velocity is calculated to be $\bar{v} \approx 0.15 \pm 0.01 \text{ m/s}$, which is ≈ 40% less than the skyrmion moving in the interior of the sample, in contrast to theoretical predictions suggesting an increased velocity at the edge. This indicates a significant pinning effect from disorder at the device edge possibly created during the lithography process.

In summary, we have observed the directional movement of magnetic skyrmions at room temperature, varying from the creep motion to the steady flow motion regime. By changing the sign of the topological charge, and the sign of the electric current, we have revealed a strong similarity between the conventional Hall effect of the electronic charge and the Hall effect due to the topological charge. Furthermore, our results suggest the important role of defects for understanding the detailed dynamics of magnetic skyrmions. In the future, similar to the well-studied motion of superconducting vortex in the presence of pinning[35], by tailoring the geometry/distribution of materials defects or artificially created pinning sites, it will be possible to experimentally reveal many exciting phenomena such as dynamic phase transitions, rectifying motion of skyrmions from ratchets, and quantized transport of magnetic skyrmions[19,31-33]. Our observations also indicate that the topological charges of magnetic skyrmions, in combination with the current induced spin Hall spin torque, can be potentially integrated for realizing novel functionalities, such as topological sorting.

**Methods**

The Ta(50Å)/Co$_{20}$Fe$_{60}$B$_{20}$(CoFeB)(11Å)/TaO$_x$(30Å) trilayer was grown onto a semi-insulating Si substrate with 300nm thick thermally formed SiO$_2$ layer by using a dc magnetron sputtering technique. TaO$_x$ layer was prepared by oxidizing the top Ta layer via oxygen plasma of 10 W for 60 s. The trilayers were annealed in vacuum for 30 minutes to induce perpendicular magnetic anisotropy. Devices were patterned by using standard photolithography and subsequent Ar ion milling. Differential polar magneto-optical Kerr effect imaging experiments were performed in a commercial MOKE microscope from Evico Magnetics. Electrical pulses were generated by using a pulse generator model 9045 from Quantum Composer.

**Acknowledgements**

Work carried out at the Argonne National Laboratory including lithographic processing and MOKE imaging was supported by the U.S. Department of Energy, Office of Science,

Materials Science and Engineering Division. Lithography was carried out at the Center for Nanoscale Materials, which is supported by the DOE, Office of Science, Basic Energy Science under Contract No. DE-AC02-06CH11357. Thin film growth performed at UCLA was partially supported by the NSF Nanosystems Engineering Research Center for Translational Applications of Nanoscale Multiferroic Systems (TANMS). Y.Z. acknowledges the support by National Natural Science Foundation of China (Project No. 1157040329), the Seed Funding Program for Basic Research and Seed Funding Program for Applied Research from the HKU, ITF Tier 3 funding (ITS/171/13 and ITS/203/14), the RGC-GRF under Grant HKU 17210014, and University Grants Committee of Hong Kong (Contract No. AoE/P-04/08). X.Z. was supported by JSPS RONPAKU (Dissertation Ph.D.) Program.

**Author contributions**

W.J., A.H. and S.G.E.te V. conceived and designed the experiments. G.Y. and K.W. fabricated the thin film. W.J., W.Z., M.B.J., and J.P. performed lithographic processing. X.Z., Y.Z. and O.H. performed micromagnetic simulation. W.J. performed MOKE experiments and data analysis. W.J., A.H. and S.G.E.te V. wrote the manuscript. All authors commented on the manuscript.

**Additional information.**

Supplementary information is available in the online version of the paper. Preprints and permission information is available online at www.nature.com/reprints. Correspondence and requests for materials should be addressed to W.J., A.H. and S.G.E.te.V.

**Competing financial interests**

Authors declare no competing financial interests.

**Figure captions:**

**Figure 1. Schematic of Hall effects for electronic and topological charges**. (**a**) Spin texture of a Néel skyrmion. (**b**) Micromagnetic simulation study of the motion of a Néel skyrmion driven by a positive electron current from left to right ($+j_e$). After applying a current, skyrmion gains a transverse velocity ($v_y$) with respect to the driving current direction ($+x$). Parameters used in this simulation are: electron current density $j_e = +3 \times 10^9$ A/m$^2$, exchange stiffness $A = 20 \times 10^{-12}$ J/m, DMI strength $D_{dmi} = 0.5$ mJ/m$^2$, perpendicular magnetic anisotropy, $K = 0.28 \times 10^6$ J/m$^3$, $M_s = 6.5 \times 10^5$ A/m, damping coefficient $\alpha = 0.02$. Blue color corresponds to magnetization orientation downwards into the plane ($-m_z$), Gray color corresponds to magnetization orientation upwards out of the plane ($+m_z$). (**c**) and (**d**) Comparison between the electronic Hall effect and skyrmion Hall effect. (**c**) For the electronic Hall effect, holes with a unit electronic charge of $+e$ accumulate at the opposite edge of device upon reversal of magnetic field directions. (**d**) For skyrmion Hall effect, the reversal of magnetic field directions (from positive to negative) reverses the sign of topological charge from $Q = -1$ (upper panel) to $Q = +1$ (lower panel), leading to the accumulation of skyrmions of topological charges at opposite edges of device.

**Figure 2. MOKE microscopy images of pulse current driven skyrmion motion.** All experiments were done by using 50 μs pulse current. (**a**) - (**e**) Snapshots of ($Q = -1$) skyrmion motion captured after applying successive current pulses of amplitude $j_e = +1.3 \times 10^6 \text{ A/cm}^2$. The skyrmion trajectory summarized in (**f**), shows no net transverse motion along the $y$ direction. (**g**) - (**k**) Snapshots of ($Q = +1$) skyrmion motion at $j_e = +1.3 \times 10^6 \text{ A/cm}^2$. The stochastic trajectory is shown in (**l**), again showing no net transverse motion. (**m**) - (**q**) Snapshots of ($Q = -1$) skyrmion motion at $j_e = +2.8 \times 10^6 \text{ A/cm}^2$. Its nearly straight and diagonal trajectory is shown in (**r**), indicating the presence of transverse motion along $+y$ direction. Two other skyrmions that moved into the frame, marked as blue circle, were not studied. (**s**) - (**w**) Snapshots of ($Q = +1$) skyrmion motion at $j_e = +2.8 \times 10^6 \text{ A/cm}^2$. Again, there is a nearly straight and diagonal trajectory shown in (**x**). However, the slope is opposite, indicating the presence of transverse motion of opposite direction (along $-y$ direction).

**Figure 3. Summary of current driven skyrmion motion.** (**a**) The average skyrmion velocity ($\bar{v}$) as a function of electron current density ($j_e$). The light blue color corresponds to the skyrmion pining regime [when $j_e < (0.6 \pm 0.1) \times 10^6 \text{ A/cm}^2$]. Light orange corresponds to the regime of stochastic motion without net transverse motion [$(0.6 \pm 0.1) \times 10^6 \text{A/cm}^2 < j_e < 1.5 \times 10^6 \text{ A/cm}^2$]. (**b**) The evolution of the skyrmion Hall angle ($\Phi_{sk}$) and the ratio between transverse and longitudinal velocities of the skyrmion ($v_y/v_x$). Light green corresponds to the regime without transverse motion

(when $j_e < 1.5 \times 10^6$ A/cm$^2$ ). When $j_e > 1.5 \times 10^6$ A/cm$^2$, both $\Phi_{sk}$ and $v_y/v_x$ are monotonically increasing as a function of current density.

**Figure 4. Transport features of skyrmions along the device edge.** (**s**) - (**w**) Snapshots of a ($Q = +1$) skyrmion moves close to the edge of the device at the same current density $j_e = +2.8 \times 10^6$ A/cm$^2$ with the same duration 50 μs as Figs. 2 (**s**) - (**w**). The skyrmion marked by the green circled is annihilated by local structural defects. The evolution of trajectory as a function of pulse number is summarized in Fig. 4 (**g**). The "oscillatory" feature indicates the absence of a well-defined skyrmion Hall angle.

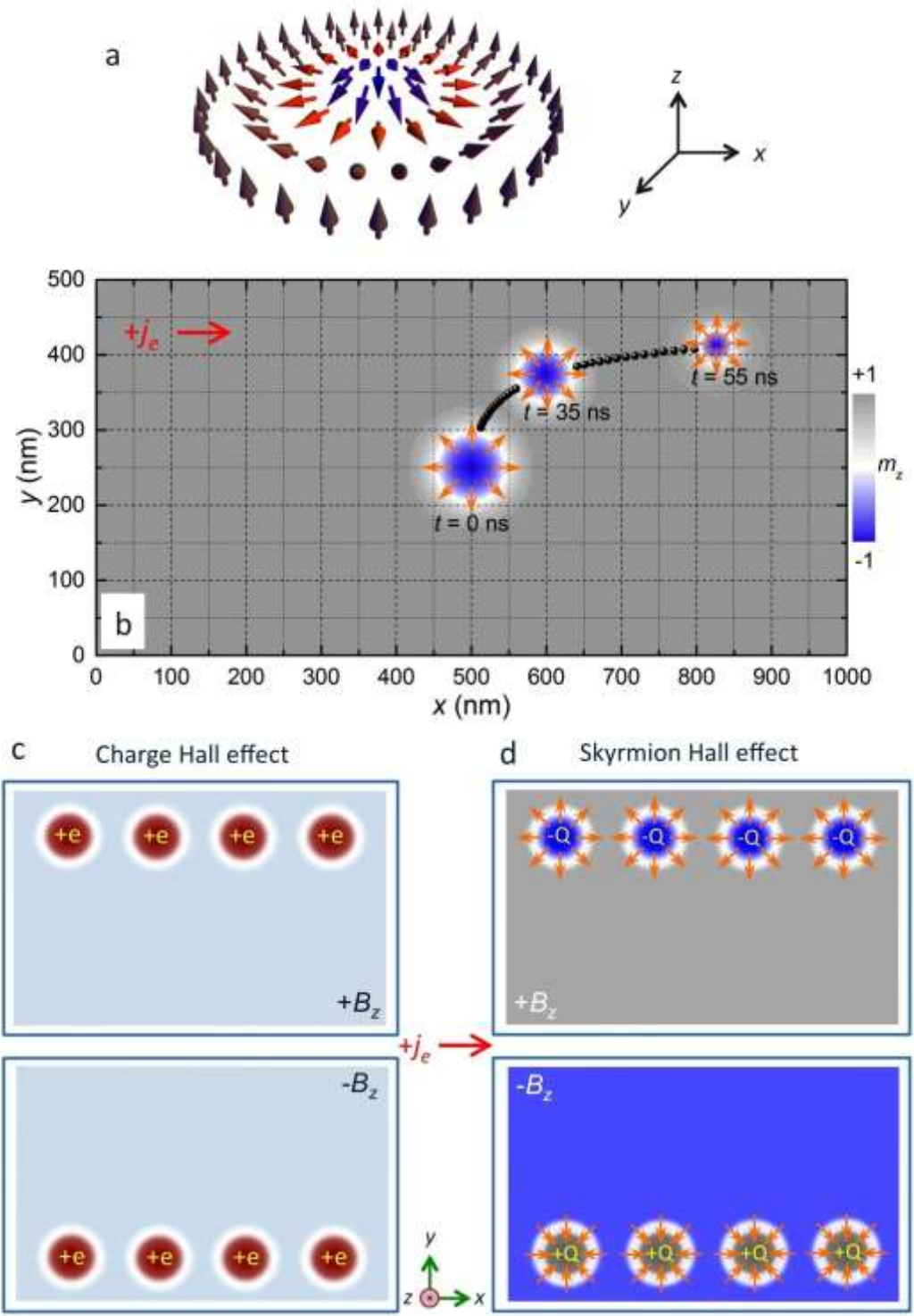

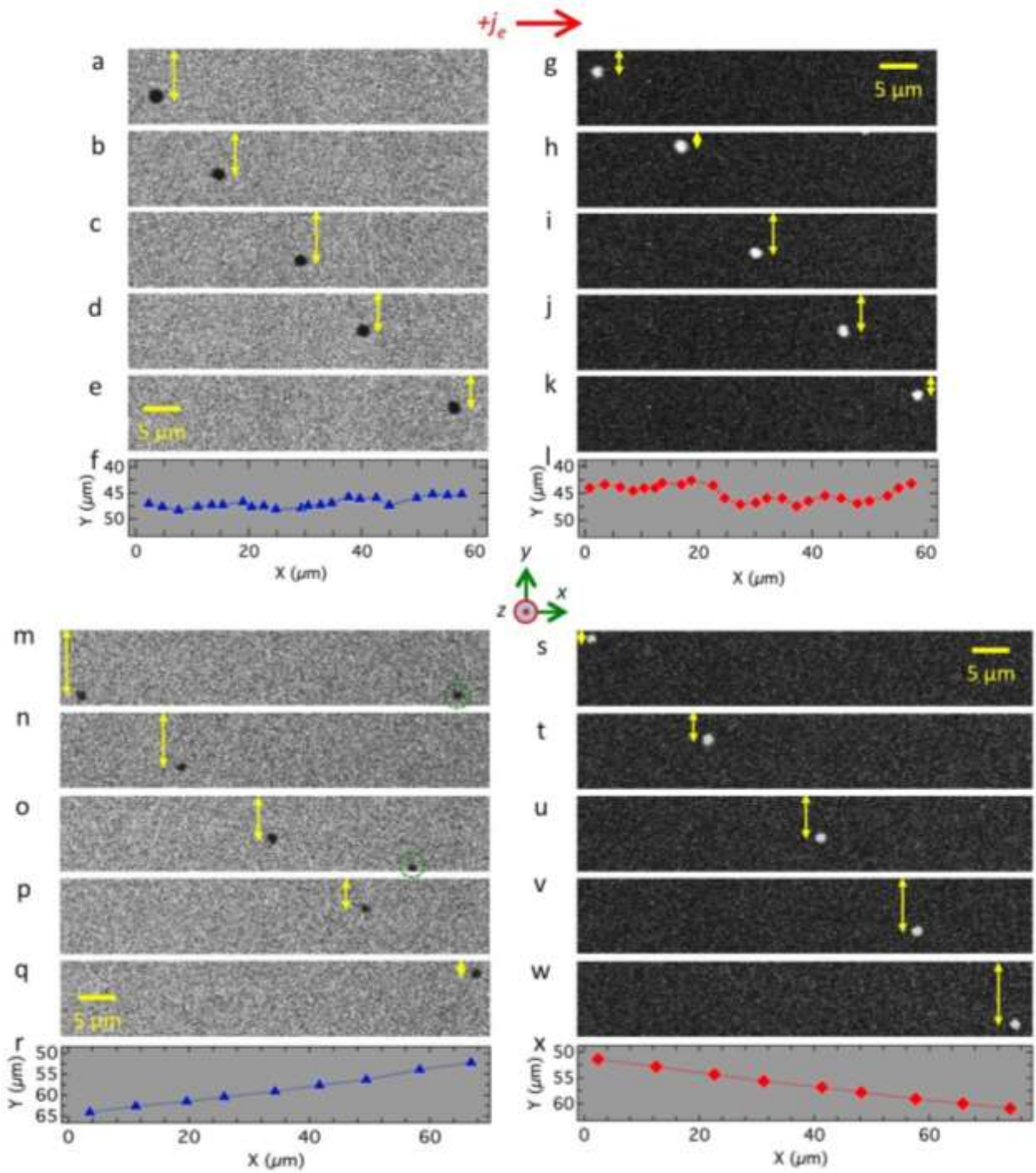

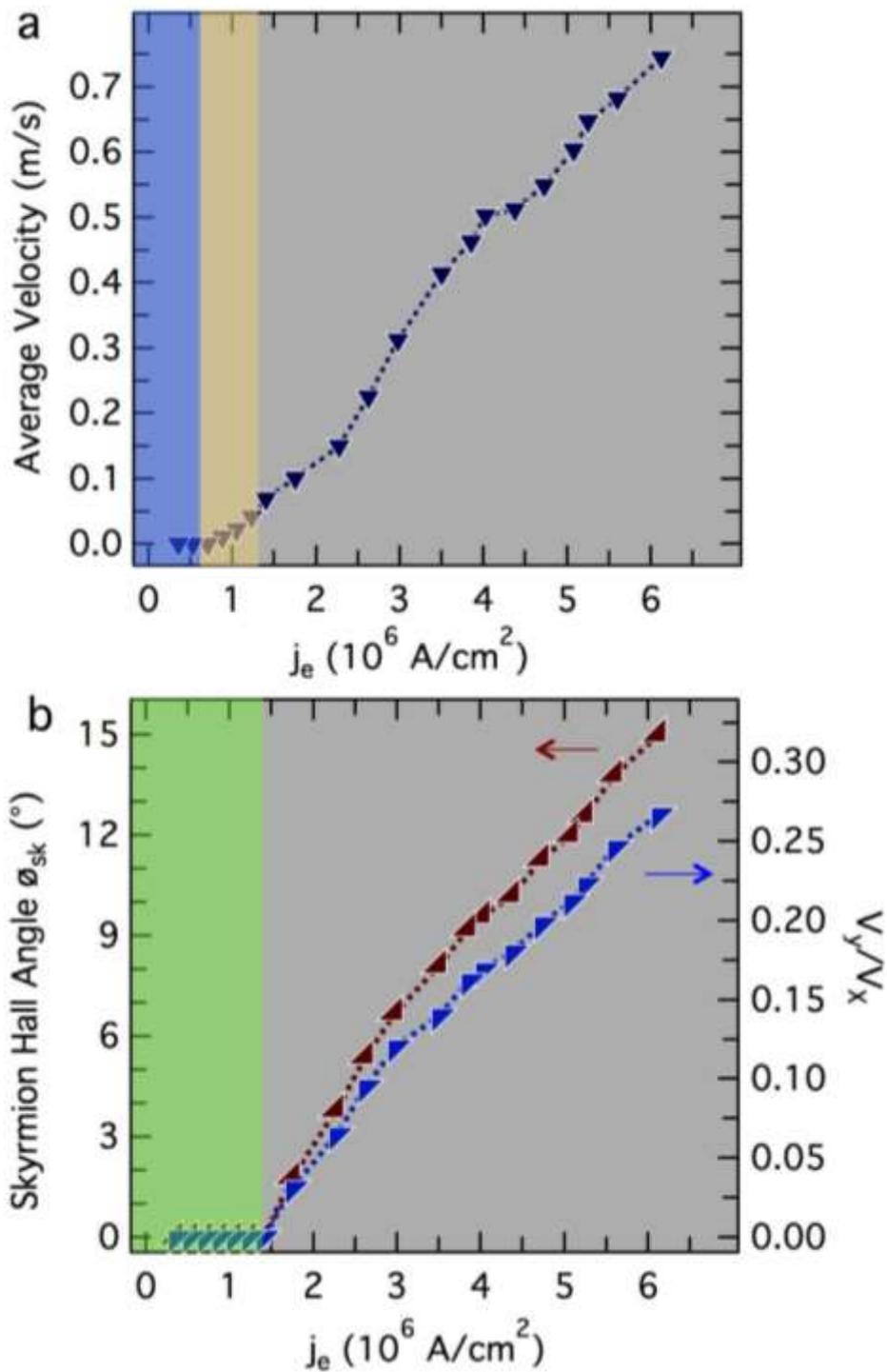

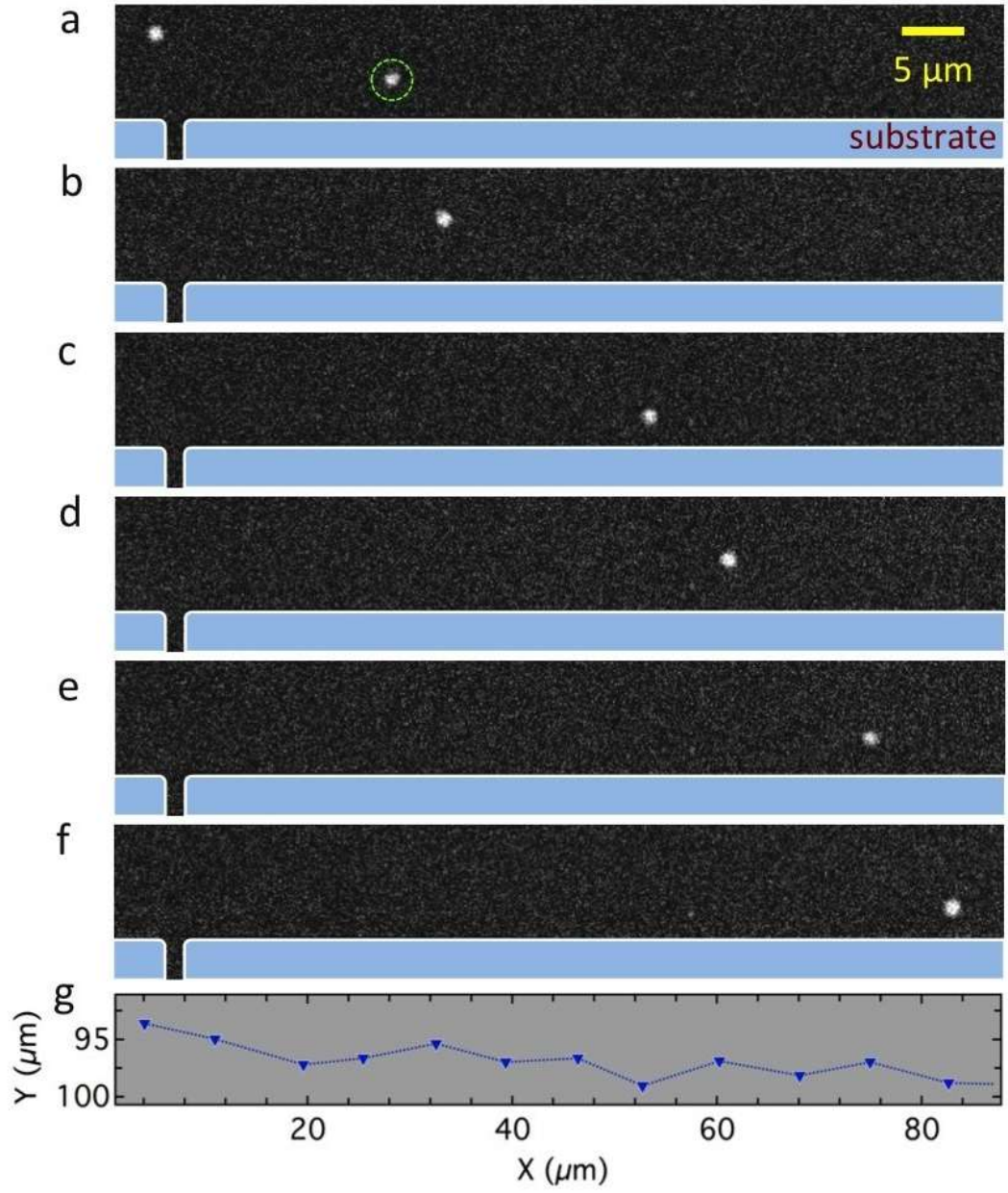

Fig. 4. Jiang et al.